\documentclass[reprint,amsmath,amssymb,aps,nofootinbib,onecolumn]{revtex4-1}                     
\usepackage{graphicx}
\usepackage{amssymb}
\usepackage{graphicx}
\usepackage{dcolumn}
\usepackage{bm}
\usepackage{xcolor}

\begin{document}

\title{Supersymmetric Dirac-Hamiltonians in $(1+1)$ dimensions revisited
}

\author{Georg Junker}
 \email{ gjunker@eso.org}
\affiliation{%
 European Southern Observatory, Karl-Schwarzschild-Stra{\ss}e 2, D-85748 Garching, Germany
}%


\date{\today}

\begin{abstract}
The most general Dirac Hamiltonians in $(1+1)$ dimensions are revisited under the requirement to exhibit a supersymmetric structure. It is found that supersymmetry allows either for a scalar or a pseudo-scalar potential. Their spectral properties are shown to be represented by those of the associated non-relativistic Witten model. The general discussion is extended to include the corresponding relativistic and non-relativistic resolvents. As example the well-studied relativistic Dirac oscillator is considered and the associated resolved kernel is found in a closed form expression by utilising the energy-dependent Green's function of the non-relativistic harmonic oscillator. The supersymmetric quasi-classical approximation for the Witten model is extended to the associated relativistic model.
\end{abstract}
%
\keywords{Dirac Equation, Green's Function, Supersymmetry, Quasi-Classical Approximation}
%
\maketitle

\section{Introduction}
\label{intro}
Dirac's well-known equation \cite{Dirac1928a,Dirac1928b} characterises the relativistic dynamics of spin-$\frac{1}{2}$ particles in the framework of quantum mechanics respecting also the principles of special relativity \cite{TH1992}. This equation has been very successful in its early days by providing a clear formalism for the spin of an electron as a point particle and also has led to the postulation of the existence of its anti-particle which was discovered shortly afterwards reassuring Dirac's interpretation \cite{Webb2008}. The Dirac equation also paved the way for quantum electrodynamics \cite{BD1} and quantum field theory of electromagnetic interactions \cite{BD2}. Nowadays, the Dirac equation is also an important tool for the description of the dynamics of charge carriers in carbon nano-structures like graphene \cite{Vafek2014} or more general in so-called Dirac electronic systems \cite{Songetal2017}. Hence exact solutions of the Dirac equation are of great interest but are rare compared to its non-relativistic counterpart, the Schr\"odinger equation. Here various techniques for finding exact solutions like the factorisation methods have been developed during the last century and more recently received much attention in the context supersymmetric methods \cite{Cooper1995,Jun2019,Cooper2001,Bagchi2001}. Also approximation methods like the WKB approach or perturbation methods are nowadays well-established for the non-relativistic quantum mechanics.

Supersymmetric (SUSY) quantum mechanics in the context of relativistic dynamics characterised by the Dirac equation was originally studied by Jackiw \cite{Jackiw1984} and Ui \cite{Ui1984}. See also the work by Cooper et al \cite{Cooper1988} and the book by Thaller \cite{TH1992}. Due to their relevance to the current paper, let us also mention more recent studies focusing on the two-dimensional graphene in a magnetic field \cite{Kuru2009,Fern14,Fern20} and those related to $(1+1)$ dimensional Dirac electron interacting with a scalar or pseudo-scalar potential \cite{Nogami1993,Nogami1997,Toyama1997,Nieto2003,Jakubsky2011,Jakubsky2012}.

As the Dirac Hamiltonian with supersymmetry is known to be closely related to a non-relativistic Pauli-Schr\"odinger type Hamiltonian \cite{Jackiw1984,Ui1984,Jun2019}, it is natural to investigate the possibility to derive exact solutions as well as approximations for a Dirac system by reducing this problem to a non-relativistic system. Whereas in the recent work \cite{JunIno2018} we focused on $(3+1)$-dimensional systems, we will limit ourselves in the current paper to supersymmetric Dirac Hamiltonians in $(1+1)$ dimensions.

We will show in the next section that supersymmetry puts a very strict condition on the most general Dirac Hamiltonian in $(1+1)$ dimensions. In essence, we find that SUSY basically allows only for either a scalar or a pseudo-scalar potential. In both cases the associated non-relativistic system is the well-studied Witten model of supersymmetric quantum mechanics \cite{Wit1981}. After recalling in section \ref{sec:3} some basic properties of supersymmetric Dirac Hamiltonians and the Witten model, we show in section \ref{sec:4} how the eigenvalue problem of the Dirac system can be reduced to that of the Witten model. As an explicit example the Dirac oscillator is considered and its spectral properties are derived from those of the standard harmonic oscillator on the real line. In section \ref{sec:5} we will study the resolvent of the Dirac system and show how this can be expressed in terms of the resolvent of the non-relativistic Witten model. Again the Dirac oscillator is chosen as an explicit example and the corresponding Green's function is derived in closed form. In section \ref{sec:6}  we will utilise the quasi-classical approximation of the Witten model to arrive at quasi-classical approximation of supersymmetric Dirac Hamiltonians. This approximation is known to respect the spectral symmetry implied by supersymmetry and this property is also respected by the derived approximation formulas of the Dirac system.  Finally in the conclusions we give an outlook how that current approach can, for example, be applied to radial Dirac Hamiltonians and a Dirac particle in a box.

\section{General Dirac Hamiltonian with Supersymmetry}
\label{sec:2}

The most general $(1+1)$-dimensional Dirac Hamiltonian, acting on Hilbert space ${\cal H} = L^2(\mathbb{R})\otimes\mathbb{C}^2$, can be put into the form
\begin{equation}\label{1}
  H_{\rm D} := c p \sigma_1 + W(x) \sigma_2 + \left[mc^2 + S(x)\right] \sigma_3 + eV(x){\bf 1},
\end{equation}
where $x$ and $p=(\hbar/{\rm i})\partial_x$ are the position and momentum operators on $L^2(\mathbb{R})$, respectively, $\{\sigma_i|i=1,2,3\}$ are the Pauli matrices and ${\bf 1}$ is the $2\times 2$ unit matrix acting on $\mathbb{C}^2$. In the above $m>0$ and $e$ stand for the mass and the charge of the Dirac particle moving along the real line $\mathbb{R}$, respectively,  and $c>0$ represents the speed of light. This particle interacts with various potentials, namely a scalar potential $S$, a pseudo-scalar potential $W$ (in $3+1$ dimensional models this is also called a tensor potential) and an electro-static potential $V$, which is the $0$-component of an electromagnetic vector potential $(V,U)$. The $1$-component $U$ of this vector potential has been omitted as it can be gauged away due the local $u(1)$ gauge-invariance. In essence, in $(1+1)$ dimensions the presence of a vector potential would lead to a pure phase factor of the form $\exp\{\pm{\rm i}(e/c)\int{\rm d}x\,U(x)\}$ in the upper and lower component of the 2-spinor, respectively.

Representing the Pauli matrices in the standard form
\begin{equation}\label{2}
  \sigma_1 = \left(\begin{array}{cc}
                     0 & 1 \\
                     1 & 0
                   \end{array}\right)\,,\qquad
  \sigma_2 = \left(\begin{array}{cc}
                     0 & -{\rm i} \\
                     {\rm i} & 0
                   \end{array}\right)\,,\qquad
  \sigma_3 = \left(\begin{array}{cc}
                     1 & 0 \\
                     0 & -1
                   \end{array}\right)\,,
\end{equation}
the above Dirac Hamiltonian can be put into a matrix form
\begin{equation}\label{3}
  H_{\rm D} = \left(\begin{array}{cc}
                      M_+& A \\
                      A^\dag & -M_-
                    \end{array}\right)\,,
\end{equation}
where
\begin{equation}\label{4}
  M_\pm := mc^2 + S(x) \pm eV(x)\,\qquad {\rm and}\qquad A:= cp-{\rm i} W(x)\,.
\end{equation}
In order to represent a supersymmetric Dirac Hamiltonian these operators are required to obey the condition \cite{TH1992,Jun2019}
\begin{equation}\label{5}
  AM_-=M_+A\,.
\end{equation}
This condition in essence guarantees that the squared Hamiltonian $H^2_{\rm D}$ becomes block diagonal and also assures the existence of an exact Foldy-Wouthuy\-sen transformation \cite{Erik1958,Beckers1990}. Above condition, however, turns out to be very restrictive and leads to constraints on the potentials with which the Dirac particle interacts. To be more explicit the constraints are $V(x) = 0$ and $S(x)=const.$, that is, the electro-static potential must vanish and the scalar potential has to be constant. Without loss of generality we also set $S=0$ as a constant $S$ can always be absorbed by the mass term $mc^2$.

Before studying the properties of such a supersymmetric Dirac Hamiltonian, let us briefly consider here alternative representations of the general Hamiltonian (\ref{1}). In fact, the representation (\ref{2}) of the Pauli matrices is the one where $\sigma_3$ is in diagonal form. Obviously, one could choose a basis in the $\mathbb{C}^2$ subspace where one of the other two Pauli matrices becomes diagonal. This would be equivalent to a cyclic permutation of the three Pauli matrices in (\ref{1}) and keeping the representation (\ref{2}) fixed. Note that the algebra obeyed by the Pauli matrices is invariant under such cyclic permutations. Hence, an alternative to (\ref{1}) would be the general Hamiltonian
\begin{equation}\label{1b}
  \tilde{H}_{\rm D} := c p \sigma_2 + W(x) \sigma_3 + \left[mc^2 + S(x)\right] \sigma_1 + eV(x){\bf 1}.
\end{equation}
Here we use a tilde to distinguish this representation for the standard one. In the matrix representation (\ref{2}) this Hamiltonian reads
\begin{equation}\label{3b}
  \tilde{H}_{\rm D} = \left(\begin{array}{cc}
                      \tilde{M}_+& \tilde{A} \\
                      \tilde{A}^\dag & -\tilde{M}_-
                    \end{array}\right)\,.
\end{equation}
Hence, we may identify the operators $\tilde{M}^\pm$ and $\tilde{A}$ as follows.
\begin{equation}\label{4b}
  \tilde{M}_\pm := eV(x)\pm W(x)\,,\qquad \tilde{A}:= -{\rm i} cp + mc^2 + S(x)\,.
\end{equation}
and the SUSY condition (\ref{5}) leads us to the restriction $W(x)=0$ and $V(x)=const$. As the constant $V$ in essence is a constant shift in the energy scale we can set it to zero without loss of generality. Note that this corresponds to the supersymmetric representation of the Dirac Hamiltonian with a scalar field as discussed by Thaller \cite{TH1992} in his section 5.5.1. See also section 9.4 in ref.\ \cite{Jun2019} and the detailed discussion by Nogami and Toyama \cite{Nogami1993}. Here we note that the case of a scalar potential can be reduced to the previous case of a pseudo-scalar potential by identifying $W(x) = mc^2 + S(x)$ and $M_\pm=0$. See, for example, ref.\ \cite{Nieto2003}. We conclude that there  is no need for a separate discussion of this case in the context of the current work.

In principle there is a third representation (we use a hat to indicate this case) of the general Hamiltonian (\ref{1}) given by
\begin{equation}\label{1c}
  \hat{H}_{\rm D} := c p \sigma_3 + W(x) \sigma_1 + \left[mc^2 + S(x)\right] \sigma_2 + eV(x){\bf 1}\,,
\end{equation}
which leads to the identification
\begin{equation}\label{4c}
  \hat{M}_\pm := eV(x)\pm cp\,\qquad {\rm and}\qquad \hat{A}:= W(x)-{\rm i} (mc^2 + S(x))\,.
\end{equation}
Here the SUSY condition (\ref{5}) implies $mc^2 + S(x)=0$ or $S(x)=-mc^2=const$. In addition we have $W(x)=const.$ Both imply that the operator $\hat{A}=W=const.$, leading to a trivial SUSY structure as will become obvious in the next section. Hence, we will not pursue this case either. That is, from now on we will, without loss of generality, limit our discussion to the case of a pure pseudo-scalar $W$, which in turn is directly applicable to the second non-trivial case of a pure scalar potential.

\section{Properties of supersymmetric Dirac Hamiltonians}
\label{sec:3}
As explicitly discussed in the previous section, without loss of generality, the most general supersymmetric Dirac Hamiltonian in $(1+1)$ dimensions can be put into the form
\begin{equation}\label{6}
  H_{\rm D}=\left(\begin{array}{cc}
                    mc^2 & cp-{\rm i} W(x)\\
                    cp +{\rm i} W(x) & -mc^2
                  \end{array}\right)\,.
\end{equation}
That is, it is of the supersymmetric form (\ref{3}) with
\begin{equation}\label{7}
  M_\pm = mc^2\,\qquad{\rm and }\qquad A=cp-{\rm i} W(x)\,.
\end{equation}
The $N=2$ SUSY structure becomes explicit by introducing the supersymmetric Hamiltonian
\begin{equation}\label{8}
  H_{\rm SUSY}:=\frac{1}{2mc^2}\left(H_{\rm D}^2-m^2c^4\right)=\left(\begin{array}{cc}
                                                                         H_+ & 0 \\
                                                                         0 & H_-
                                                                       \end{array}\right)
\end{equation}
with partner Hamiltonians
\begin{equation}\label{9}
  H_+ := \frac{1}{2mc^2}AA^\dag\,,\qquad H_- := \frac{1}{2mc^2}A^\dag A
\end{equation}
and the SUSY charges
\begin{equation}\label{10}
  Q:= \frac{1}{2mc^2}\left(\begin{array}{cc} 0 & A \\ 0 & 0 \end{array}\right)\,,\qquad
  Q^\dag = \frac{1}{2mc^2}\left(\begin{array}{cc} 0 & 0 \\ A^\dag & 0 \end{array}\right)\,.
\end{equation}
These operators obey the $N=2$ SUSY algebra
\begin{equation}\label{11}
  H_{\rm SUSY}=\{Q,Q^\dag\}\,,\qquad Q^2 = 0 =(Q^\dag)^2\,,\qquad \{Q,\sigma_3\}=0\,,
  \qquad [ H_{\rm SUSY}, \sigma_3 ] = 0 \,,\qquad \sigma_3^2 = 1\,,
\end{equation}
where $\sigma_3$ plays the role of the grading (or Witten) operator. In fact, the partner Hamiltonians have the explicit form
\begin{equation}\label{12}
  H_\pm =\frac{p^2}{2m} +\Phi^2(x)\pm \frac{\hbar}{\sqrt{2m}}\Phi'(x)\,,
\end{equation}
which is the well-known one-dimensional non-relativistic Witten model of supersymmetric quantum mechanics \cite{Wit1981,Jun2019}. Note that in the above we have rescaled the pseudo-scalar potential $W(x)=: \sqrt{2mc^2}\, \Phi(x)$, where $\Phi$ may now be identified with the SUSY potential of the Witten model.

This Witten model has been studied extensively in the last 25 years and finds many applications in various fields of physics. Here let us summarise the most essential properties of this model. Obviously, both partner Hamiltonians are non-negative and furthermore are essential isospectral, that means, there strictly positive eigenvalues are identical. Let us assume that both have a pure discrete spectrum represented by the eigenvalues $\varepsilon_n$ with associated eigenstates $\phi_n^\pm$. That is, we have
\begin{equation}\label{13}
  H_\pm \phi^\pm_n = \varepsilon_n\phi^\pm_n \,,\qquad \varepsilon_n >0 \,,\qquad n=1,2,3,\ldots\,.
\end{equation}
In addition, in case of an unbroken SUSY, there exists a zero-energy ground state, which for convenience is then assumed to belong to $H_-$. This can always be achieved with a change of sign in the SUSY potential. Hence, in the case of unbroken SUSY, we have in addition the ground state $\phi^-_0$ associated with $\varepsilon_0=0$ defined via $A\phi^-_0=0$ leading to the explicit form
\begin{equation}\label{14}
  \phi^-_0(x) = N\exp\left\{-\frac{\sqrt{2m}}{\hbar}\int {\rm d} x\,\Phi(x)\right\} = N\exp\left\{-\frac{1}{\hbar c}\int {\rm d} x\,W(x)\right\}
\end{equation}
with $N$ denoting a normalisation constant.
The eigenstates (\ref{13}) belonging to the strictly positive spectrum are related to each other via the SUSY transformations
\begin{equation}\label{15}
  A\phi^-_n = \sqrt{2mc^2\varepsilon_n}\phi^+_n\,,\qquad A^\dag\phi^+_n = \sqrt{2mc^2\varepsilon_n}\phi^-_n\,,\qquad n=1,2,3,\ldots\,.
\end{equation}

Before studying the spectral properties of $H_{\rm D}$ let us briefly comment on the supersymmetric representation which covers the case of a scalar potential $S$.  As pointed out in the previous section, the case of a pure scalar potential can be covered by above discussion when we identify
\begin{equation}\label{scalarcase}
  W(x) := mc^2 + S(x)\,,\quad \tilde{A} := -{\rm i} \left(cp+{\rm i}W(x)\right)\,,\quad H_{\rm SUSY} = (\tilde{H}^2_{\rm D}-e^2V^2)/2mc^2\,,
\end{equation}
where $V=const.$

It is now also obvious that the third case mentioned in section \ref{sec:3} does lead to a trivial SUSY structure as in this case $\hat{A}=W=const.$, which leads to trivial partner Hamiltonians $H_\pm = W^2/2mc^2=const.$

\section{Spectral properties of the Hamiltonian}
\label{sec:4}
In this section we will review the spectral properties of the most general supersymmetric Dirac Hamiltonian in $(1+1)$ dimensions, which is fully characterised by a pseudo-scalar potential as shown in the previous section, cf.\ eq.\ (\ref{6}).
Such supersymmetric Dirac Hamiltonians are known to be block-diagonalisable via a unitary transformation separating positive and negative energy eigenspaces. Indeed, it is possible to show that there exists a unitary matrix $U$, see for example ref.\ \cite{TH1992,Jun2019}, which transforms the Dirac Hamiltonian (\ref{6}) into the form
\begin{equation}\label{3.1}
  H_{\rm FW}:= U H_{\rm D} U^\dag =\left(\begin{array}{cc}
                                                  \sqrt{2mc^2H_++m^2c^4} & 0 \\
                                                  0 & -\sqrt{2mc^2H_-+m^2c^4}
                                                \end{array}\right)\,.
\end{equation}
Hence the positive and negative energy spectrum of $H_{\rm FW}$ and hence also of $H_{\rm D}$ is fully determined by that of $H_+$ and $H_-$, respectively. As both partner Hamiltonians are essential isospectral, the spectrum of $H_{\rm D}$ is in fact symmetric about the origin,
\begin{equation}\label{3.2}
  E^\pm_n = \pm\sqrt{2mc^2\varepsilon_n + m^2c^4}\,,\qquad n>0\,.
\end{equation}
In addition, in the case of an unbroken SUSY the eigenvalue $E^-_0=-mc^2$ also belongs to the spectrum of $H_{\rm D}$. The corresponding eigenstates are given by
\begin{equation}\label{3.3}
  \psi_n^\pm = U^\dag \tilde{\psi}^\pm_n\,,\qquad{\rm with}\qquad
  \tilde{\psi}^+_n=\left(\begin{array}{c}\phi_n^+ \\ 0 \end{array} \right)\,,\qquad
  \tilde{\psi}^-_n=\left(\begin{array}{c}0 \\ \phi_n^- \end{array} \right)\,.
\end{equation}
In other words the spectral properties of $H_{\rm D}$, that is,
\begin{equation}\label{3.4}
  H_{\rm D} \psi^\pm_n = E^\pm _n\psi^\pm_n\,,
\end{equation}
are fully characterised by those of $H_{\rm SUSY}$. In case of unbroken SUSY in addition we have
\begin{equation}\label{3.5}
  H_{\rm D} \psi^-_0 = -mc^2\psi^-_0\,,\qquad \psi^-_0 = U^\dag\left(\begin{array}{c}0 \\ \phi_0^- \end{array} \right)
  = \left(\begin{array}{c}0 \\ \phi_0^- \end{array} \right)\,.
\end{equation}
Note that $U=1$ on $\ker H_-$ as will be shown below.

Let us now study the unitary transformation matrix $U$ in more detail. According to the general SUSY approach \cite{TH1992,Jun2019} this matrix is given by
\begin{equation}\label{3.6}
  U := a_+ + \sigma_3 \,{\rm sgn}\, Q_1 \, a_-\, \qquad {\rm with} \qquad a_\pm := \sqrt{\frac{|H_{\rm D}|\pm mc^2}{2 |H_{\rm D}|}}\,.
\end{equation}
Here the self-adjoint supercharge $Q_1$ is defined as follows
\begin{equation}\label{3.7}
  Q_1 := \sqrt{2mc^2}\left(Q+Q^\dag\right)=\left(\begin{array}{cc}0 & A \\ A^\dag & 0 \end{array}\right)
\end{equation}
and the definition of the sign function is
\begin{equation}
  {\rm sgn}\,x := \left\{\begin{array}{c} +1 \\ 0 \\ -1 \end{array}\right. \qquad {\rm for} \qquad
                         \begin{array}{c} x > 0 \\ x = 0 \\ x <0 \end{array} \,.
\end{equation}
To make things a bit more explicit let us first look into the spectral properties of $Q_1$. It is straight forward to show with the help of the SUSY transformations (\ref{15}) that the states
\begin{equation}\label{3.8}
  \chi_n^\pm := \frac{1}{\sqrt{2}}\left(\begin{array}{c}\phi^+_n \\ \pm \phi^-_n \end{array}\right)
\end{equation}
are eigenstates of $Q_1$, that is,
\begin{equation}\label{3.9}
  Q_1\chi_n^\pm = \pm \sqrt{2mc^2\varepsilon_n}\chi_n^\pm\,.
\end{equation}
In addition, for unbroken SUSY, $Q_1$ has a zero eigenvalue as
\begin{equation}\label{3.10}
  Q_1\chi_0^- = 0\qquad{\rm for}\qquad\chi_0^- := \left(\begin{array}{c}0\\ \phi^-_0 \end{array}\right)\,,
\end{equation}
which implies $U= {\bf 1}$ on ${\rm ker}\, H_-={\rm span}\,|\phi_0^-\rangle\langle\phi_0^-|$.
Having this in mind it is obvious that the operator ${\rm sgn}\,Q_1$ can explicitly be written as
\begin{equation}\label{3.11}
  {\rm sgn}\,Q_1 = \left(\begin{array}{cc} 0 & \left(AA^\dag\right)^{-1/2}A \\ \left(A^\dag A\right)^{-1/2}A^\dag & 0 \end{array}\right)\,.
\end{equation}
Note that ${\rm sgn}\,Q_1\chi_n^\pm=\pm\chi_n^\pm$.

Together with the diagonal matrix
\begin{equation}\label{3.12}
  |H_{\rm D}|=\left(\begin{array}{cc} \sqrt{AA^\dag +m^2c^4} & 0 \\ 0 & \sqrt{A^\dag A+m^2c^4} \end{array}\right)
\end{equation}
and the dimensionless operator $a:=A/mc^2$ we have
\begin{equation}\label{3.13}
  a_\pm =\left(\begin{array}{cc} \displaystyle\sqrt{\frac{\sqrt{aa^\dag +1} \pm 1}{2\sqrt{aa^\dag +1}}} & 0 \\
                                 0 & \displaystyle\sqrt{\frac{\sqrt{a^\dag a + 1} \pm 1}{2\sqrt{a^\dag a + 1}}}
  \end{array}\right)
\end{equation}
leading to the explicit but rather complicated expression
\begin{equation}\label{3.14}
  U=\frac{1}{\sqrt{2}}\left(
  \begin{array}{cc}
  \displaystyle
    \sqrt{\frac{\sqrt{aa^\dag +1} + 1}{\sqrt{aa^\dag +1}}} & \displaystyle
                       (aa^\dag)^{-1/2}a \sqrt{\frac{\sqrt{a^\dag a + 1} - 1}{\sqrt{a^\dag a + 1}}}\\[6mm]
    \displaystyle
    -(a^\dag a)^{-1/2}a^\dag \sqrt{\frac{\sqrt{aa^\dag +1} - 1}{\sqrt{aa^\dag +1}}}  &\displaystyle
                        \sqrt{\frac{\sqrt{a^\dag a + 1} + 1}{\sqrt{a^\dag a + 1}}}
  \end{array}
  \right).
\end{equation}
Noting that
\begin{equation}\label{3.15}
  a\phi_n^- = \sqrt{\frac{2\varepsilon_n}{mc^2}} \phi_n^+\,,\qquad
  a^\dag\phi_n^+ = \sqrt{\frac{2\varepsilon_n}{mc^2}} \phi_n^-\,,
\end{equation}
which implies
\begin{equation}\label{3.16}
  a^\dag(aa^\dag)^{-1/2}\phi_n^+=\phi_n^-\,,\qquad a(a^\dag a)^{-1/2}\phi_n^-=\phi_n^+\,,
\end{equation}
we obtain explicit expressions for the eigenfunctions (\ref{3.3}) of the original Dirac Hamiltonian (\ref{6}) in terms of the eigenfunctions of the corresponding non-relativis\-tic Witten model
\begin{equation}\label{3.17}
\begin{array}{l}
    \psi_n^+ = 
    \displaystyle
    \frac{1}{\sqrt{2}}\left(\begin{array}{c}
       \sqrt{1+\left(1+2\varepsilon_n/mc^2\right)^{-1/2}}\phi_n^+ \\[2mm]
       \sqrt{1-\left(1+2\varepsilon_n/mc^2\right)^{-1/2}}\phi_n^-
       \end{array} \right) =
     \frac{1}{\sqrt{2}}\left(\begin{array}{c}
       \sqrt{1+\frac{mc^2}{E_n^+}}\phi_n^+ \\[2mm]
       \sqrt{1+\frac{mc^2}{E_n^-}}\phi_n^-
       \end{array} \right) ,\\[6mm]
    \psi_n^- = 
    \displaystyle
    \frac{1}{\sqrt{2}}\left(\begin{array}{c}
       - \sqrt{1-\left(1+2\varepsilon_n/mc^2\right)^{-1/2}}\phi_n^+ \\[2mm]
       \sqrt{1+\left(1+2\varepsilon_n/mc^2\right)^{-1/2}}\phi_n^-
       \end{array} \right)
        = \displaystyle
    \frac{1}{\sqrt{2}}\left(\begin{array}{c}
       - \sqrt{1-\frac{mc^2}{E_n^+}}\phi_n^+ \\[2mm]
       \sqrt{1-\frac{mc^2}{E_n^-}}\phi_n^-
       \end{array} \right),\\[6mm]
    \psi_0^- = 
               \left(\begin{array}{c} 0 \\ \phi_0^- \end{array} \right).
\end{array}
\end{equation}
Note that these states constitute a complete orthonormal set on ${\cal H}$ due to the orthonormality of $\langle \phi_m^+|\phi_n^+\rangle = \langle \phi_m^-|\phi_n^-\rangle= \delta_{mn}$ and the relation $E^-_n=-E^+_n$.
Hence, we have reduced the complete eigenvalue problem of a general supersymmetric Dirac Hamiltonian (\ref{6}) to that of the associated non-relativistic Witten model (\ref{12}). Above results cover several special cases discusses before. For example, they agree with previous work exclusively dedicated to the Dirac oscillator \cite{Toyama1997,SZM2001}.
The special case of a vanishing mass characterising a Dirac electron in graphene under the influence of a magnetic field has in essence been discussed by Kuru, Negro and Nieto \cite{Kuru2009} and agrees with above result when taking the limit $m\to 0$.

\subsection{The Dirac oscillator in $(1+1)$ dimensions}
\label{sec:4.1}
As an illustrative example let us consider the $(1+1)$-dimensional version of the Dirac oscillator \cite{Mos1989} which is  characterised by the pseudo-scalar potential
\begin{equation}\label{3.18}
  W(x) = mc\omega x\,, \qquad \omega > 0\,.
\end{equation}
Obviously the two partner Hamiltonians are represented by a shifted harmonic oscillator
\begin{equation}\label{3.19}
  H_\pm = \frac{p^2}{2m}+\frac{m}{2}\omega^2x^2\pm\frac{\hbar\omega}{2}\,.
\end{equation}
The corresponding spectral properties of these Hamiltonians are well known and given by
\begin{equation}\label{3.20}
  \varepsilon_n = \hbar\omega n\,,\qquad \phi_n^+ = |n-1\rangle\,, \qquad\phi_n^-=|n\rangle\,.
\end{equation}
Here $n\in \mathbb{N}$ for $H_+$, whereas $n\in \mathbb{N}_0$ for $H_-$, and $|n\rangle$ denotes a standard harmonic oscillator eigenstate which obeys the relations
\begin{equation}\label{3.21}
  b|n\rangle = \sqrt{n}|n-1\rangle\,,\qquad b^\dag|n\rangle = \sqrt{n+1}|n+1\rangle\,,
\end{equation}
with $b:={\rm i} p/m\omega +x$ and $b^\dag= -{\rm i} p /m\omega +x$ being the standard annihilation and creation operators of the harmonic oscillator. They are related to the dimensionless operators $a$ and $a^\dag$ introduced above as follows:
\begin{equation}\label{3.22}
  a=-{\rm i}\sqrt{\frac{2\hbar\omega}{mc^2}}\,b\,,\qquad a^\dag= {\rm i}\sqrt{\frac{2\hbar\omega}{mc^2}}\,b^\dag\,.
\end{equation}
Hence via (\ref{3.2}) and (\ref{3.17}) we immediately find that the eigenvalues and eigenstates of the Dirac oscillator Hamiltonian
\begin{equation}\label{3.23}
  H_{\rm D}=\left(\begin{array}{cc}
                    mc^2 & cp-{\rm i} mc\omega x \\ cp+ {\rm i} mc\omega x & -mc^2 \end{array}
  \right)
\end{equation}
are given by
\begin{equation}\label{3.24}
\begin{array}{l}
  E_n^+= mc^2\sqrt{1+2n\frac{\hbar\omega}{mc^2}}\,,\qquad
  \psi_n^+={\displaystyle\frac{1}{\sqrt{2}}}\left(\begin{array}{c} \sqrt{1+\left(1+2n\frac{\hbar\omega}{mc^2}\right)^{-1/2}}\,|n-1\rangle  \\[2mm]
  \sqrt{1-\left(1+2n\frac{\hbar\omega}{mc^2}\right)^{-1/2}}\,|n\rangle \end{array} \right)\,,
  \\[6mm]
  E_n^-= -mc^2\sqrt{1+2n\frac{\hbar\omega}{mc^2}}\,,\qquad
  \psi_n^-={\displaystyle\frac{1}{\sqrt{2}}}\left(\begin{array}{c} -\sqrt{1-\left(1+2n\frac{\hbar\omega}{mc^2}\right)^{-1/2}}\,|n-1\rangle  \\[2mm]
  \sqrt{1+\left(1+2n\frac{\hbar\omega}{mc^2}\right)^{-1/2}}\,|n\rangle \end{array} \right)\,,
\end{array}\end{equation}
where $n=0$ is only allowed for the second line with $\psi^-_0$ having only a lower component, cf.\ last line in eq.\ (\ref{3.17}). As mentioned before, these results are not new and agree with previous work dedicated to the Dirac oscillator in $(1+1)$ dimensions \cite{Toyama1997,SZM2001}.
Let us note that we can also recover from this the zero-mass case discussed by Kuru, Negro and Nieto \cite{Kuru2009} when setting in the above result $2m\omega \to \omega$ and taking the limit $m\to 0$.

\section{Spectral properties of the Resolvent}
\label{sec:5}
In this section we will study the resolvent of a general supersymmetric Dirac Hamiltonian (\ref{6}) defined by
\begin{equation}\label{4.1}
  G_{\rm D}(z):=\frac{1}{H_{\rm D}-z}\,,\qquad z\in\mathbb{C}\backslash {\rm spec}\, H_{\rm D}
\end{equation}
which can be expressed in terms of the so-called iterated resolvent $g$ as follows
\begin{equation}\label{4.2}
  G_{\rm D}(z) = \left(H_{\rm D} +z\right)g(z^2)\,, \qquad g(z^2):= \frac{1}{H^2_{\rm D}-z^2}\,.
\end{equation}
As $H^2_{\rm D}$ is block-diagonal,
\begin{equation}\label{4.3}
  H^2_{\rm D} =2mc^2\left(\begin{array}{cc} H_+ & 0 \\ 0 & H_- \end{array}\right)+m^2c^4\,,
\end{equation}
so is
\begin{equation}\label{4.4}
  g(z^2)=\frac{1}{2mc^2}\left(\begin{array}{cc} G_+(\zeta(z)) & 0 \\ 0 & G_-(\zeta(z)) \end{array}\right)\,,
\end{equation}
where
\begin{equation}\label{4.5}
  G_\pm(\zeta) := \frac{1}{H_\pm -\zeta}
\end{equation}
denotes the resolvent for the SUSY partner Hamiltonians (\ref{12}) and we have introduced
\begin{equation}\label{zeta}
\zeta(z):= \frac{z^2}{2mc^2}-\frac{mc^2}{2}\,.
\end{equation} Hence we can express the resolvent (\ref{4.1}) in terms of the non-relativistic resolvents (\ref{4.5}) as follows
\begin{equation}\label{4.6}
G_{\rm D}(z)=\frac{1}{2mc^2}\left(\begin{array}{cc} (z+mc^2)G_+(\zeta(z)) & AG_-(\zeta(z))\\[2mm] A^\dag G_+(\zeta(z))& (z - mc^2)G_-(\zeta(z))\end{array}\right)\,.
\end{equation}
Utilising the spectral representations
\begin{equation}\label{4.7}
  G_\pm(\zeta)=\sum_{\varepsilon_n\geq 0}\frac{|\phi_n^\pm\rangle\langle\phi_n^\pm|}{\varepsilon_n -\zeta}\,,
\end{equation}
where the $\varepsilon_0=0$ term is only present in $G_-$ in case of an unbroken SUSY, we arrive with the help of the SUSY transformations (\ref{15}) at the spectral representation of (\ref{4.1})
\begin{equation}\label{4.8}
\begin{array}{l}
  G_{\rm D}(z)=\left(\begin{array}{cc} 0 & 0\\ 0 & \displaystyle
  \frac{|\phi^-_0\rangle\langle\phi^-_0|}{z+mc^2}\end{array}\right)
   \displaystyle
   + \sum_{\varepsilon_n>0}\frac{1}{2mc^2\varepsilon_n +m^2c^4-z^2}
  \left(\begin{array}{cc} \left(z+mc^2\right)~|\phi^+_n\rangle\langle\phi^+_n|~~~&
        \left(2mc^2\varepsilon_n\right)^{1/2} |\phi^+_n\rangle\langle\phi^-_n|\\[2mm]
   \left(2mc^2\varepsilon_n\right)^{1/2} |\phi^-_n\rangle\langle\phi^+_n| ~~~ &
        \left(z-mc^2\right)~ |\phi^-_n\rangle\langle\phi^-_n|\end{array}\right).
\end{array}
\end{equation}
Again the first term, which has a pole at $E_0=-mc^2$, is only present in case of an unbroken SUSY. The poles of the second term reflect the energy eigenvalues as given in (\ref{3.2}) as expected.

Note that besides the spectral representation, the diagonal form (\ref{4.4}) also allows to derive a path integral representation of the Green's function following, in essence, the same procedure as presented in ref.\ \cite{JunIno2018}. Below we will derive the Green's function of the Dirac oscillator in closed form directly from that of the standard harmonic oscillator.

\subsection{The Green's function of the Dirac oscillator}
\label{sec:5.1}
Let us reconsider the Dirac oscillator in $(1+1)$ dimensions. As we have seen in the previous section the partner Hamiltonians of the Dirac oscillator are related to the standard harmonic oscillator Hamiltonian
\begin{equation}\label{4.9}
  H_0 :=  \frac{p^2}{2m}+\frac{m}{2}\omega^2x^2
\end{equation}
via constants shifts given by $H_\pm = H_0 \pm\hbar\omega/2$. Hence the two partner resolvents (\ref{4.7}) can be obtained from the usual harmonic oscillator resolvent
\begin{equation}\label{4.10}
  G_0(\zeta):=\frac{1}{H_0-\zeta}
\end{equation}
via the relation $G_\pm(\zeta)=G_0(\zeta\mp\hbar\omega/2)$. Following a recent work by Glasser and Nieto \cite{GlaNie2015} the coordinate representation of (\ref{4.10}) can be given in closed form
\begin{equation}\label{4.11}
\begin{array}{rl}
  G_0(x'',x';\hbar\omega\epsilon)& :=\langle x''|G_0(\hbar\omega\epsilon)|x'\rangle 
   \displaystyle =\sqrt{\frac{m}{\pi\omega\hbar^3}}\Gamma\left(\textstyle\frac{1}{2}-\epsilon\right)
  D_{\epsilon-1/2}(\mu x_+)D_{\epsilon-1/2}(-\mu x_-)\,.
\end{array}
\end{equation}
In the above $D_\nu$ stands for the parabolic cylinder function, $\Gamma$ is Euler's gamma function, $\mu:=\sqrt{2m\omega/\hbar}$ and $x_\pm$ stands for the maximum and minimum of $x''$ and $x'$, respectively, that is, $x_+:=\max (x'',x')$ and $x_-:=\min (x'',x')$. For simplicity we have introduced the dimensionless parameter $\epsilon := \zeta/\hbar\omega=\frac{z^2}{2mc^2\hbar\omega}-\frac{mc^2}{2\hbar\omega}$.
With this explicit expression it is now also possible to find the corresponding closed-form expression for the Green's function of the Dirac oscillator. In fact, with the help of the relations
\begin{equation}\label{4.12}
D'_\nu(y)+(y/2)D_\nu(y)= \nu D_{\nu-1}(y)\,,\quad -D'_\nu(y)+(y/2)D_\nu(y)= D_{\nu+1}(y)
\end{equation}
one finds following useful relations for the creation and annihilation operators (\ref{3.21}),
\begin{equation}\label{4.13}
  b D_\nu(\mu x)=\nu D_{\nu-1}(\mu x)\,,\quad b^\dag D_\nu(\mu x) = D_{\nu +1} (\mu x)\,, \quad b^\dag b D_\nu(\mu x) = \nu D_{\nu} (\mu x)\,.
\end{equation}
From these relations it is straight-forward to obtain the closed-form expressions
\begin{equation}\label{4.14}
  \begin{array}{l}
    \displaystyle
    \frac{1}{2mc^2}\langle x''|AG_-(\zeta)|x'\rangle= \frac{\rm i}{\hbar c}\frac{{\rm sgn} (x''-x')}{\sqrt{2\pi}} \Gamma(1-\epsilon)D_{\epsilon-1}(\mu x_+)D_\epsilon(-\mu x_-) \\[4mm]
    \displaystyle
    \frac{1}{2mc^2}\langle x''|A^\dag G_+(\zeta)|x'\rangle= \frac{\rm i}{\hbar c}\frac{{\rm sgn} (x''-x')}{\sqrt{2\pi}} \Gamma(1-\epsilon)D_{\epsilon}(\mu x_+)D_{\epsilon -1}(-\mu x_-)\\[4mm]
    \displaystyle
    \frac{z+mc^2}{2mc^2}\langle x''|G_+(\zeta)|x'\rangle=\frac{1}{\hbar c}\frac{z+mc^2}{\sqrt{2\pi mc^2\hbar\omega}}
    \Gamma(1-\epsilon)D_{\epsilon}(\mu x_+)D_{\epsilon}(-\mu x_-)\\[4mm]
    \displaystyle
    \frac{z-mc^2}{2mc^2}\langle x''|G_-(\zeta)|x'\rangle=\frac{1}{\hbar c}\frac{z-mc^2}{\sqrt{2\pi mc^2\hbar\omega}}
    \Gamma(-\epsilon)D_{\epsilon-1}(\mu x_+)D_{\epsilon-1}(-\mu x_-)
  \end{array}
\end{equation}
which constitute the four components of the resolvent (\ref{4.6}) in the coordinate representation, that is $\langle x''|G_{\rm D}(z)|x'\rangle$, for the Dirac oscillator (\ref{3.23}). Obviously the pole at $\epsilon = 0$ leads to the eigenvalue $z=E^-_0=-mc^2$ whereas the other poles at $\epsilon = n \in\mathbb{N}$ result in the eigenvalues as given in (\ref{3.24}).

\section{Quasi-classical approximation}
\label{sec:6}
In the previous section we have seen that the supersymmetric Dirac problem in $(1+1)$ dimensions can be reduced to the corresponding non-relativistic Witten model. That is, the spectral properties of $H_{\rm D}$ are fully deduced from those of the partner Hamiltonians $H_\pm$. In particular, all exactly solvable Witten models, e.g.\ those which are shape-invariant, immediately result in the exact solutions of the corresponding Dirac system. Even closed form expression of the resolvent $G_{\rm D}$ can be obtained when the corresponding resolvents $G_\pm$ are given in closed form.

However, the above general discussion of section \ref{sec:3} is not limited to only exactly solvable Witten models. It may also be applied to approximation methods. Here we consider the so-called quasi-classical approximations to the energy eigenvalues of the Witten model. In fact, this quasi-classical approximation is in general better than the usual WKB approximation as it respects the SUSY induced symmetry between the eigenvalues of the two partner Hamiltonians and is exact for the ground-state energy in case of unbroken SUSY \cite{Jun2019}. To be more explicit, in case of unbroken SUSY with the ground state belonging to $H_-$ the eigenvalues $\varepsilon_n$ for the SUSY partners $H_\pm$ are obtained via the so-called CBC formula\footnote{Here CBC stands for Comtet, Bandrauk and Campbell \cite{CBC1985} as these authors were the first pointing out that this modified approximation formula of the well-known WKB approximation yields the exact spectrum in case of a shape-invariant $W$ for unbroken SUSY.}
\begin{equation}\label{5.1}
  \int_{x_{\rm L}}^{x_{\rm R}}{\rm d} x \sqrt{2m\left(\varepsilon-\Phi^2(x)\right)} = \hbar\pi n\,,\qquad n\in\mathbb{N}_0\,,
\end{equation}
where the left and right turning points $x_{\rm L} \leq x_{\rm R}$ are defined by $\Phi^2(x_{\rm L/R})=\varepsilon$. Note that as before $n=0$ is only allowed for $H_-$ and obviously implies $\varepsilon_0 = 0$, which is exact. For $n\in\mathbb{N}$ above CBC formula provides a quasi-classical approximation to the jointed eigenvalues of $H_\pm$. In case of shape-invariant systems this approximation is even exact for any $n$.

In the case of a broken SUSY the corresponding quasi-classical approximation is given by the so-called EIJ formula\footnote{This version, applicable to the case of broken SUSY, is due to Eckhardt \cite{Eck1986} and Inomata and Junker \cite{IJ1993} and provides exact spectral values in case of a shape-invariant $W$ with broken SUSY. For a detailed discussion of both the CBC and EIJ formula see chapter 6 in \cite{Jun2019}, where also the approximate energy eigenfunctions are discussed to some detail.} and reads
\begin{equation}\label{5.2}
  \int_{x_{\rm L}}^{x_{\rm R}}{\rm d} x \sqrt{2m\left(\varepsilon-\Phi^2(x)\right)} = \hbar\pi \left(n-\frac{1}{2}\right)\,,\qquad n\in\mathbb{N}\,.
\end{equation}
Again this formula reflects the strict iso-spectral property of the partner Hamilton\-ians as well as the strict positivity $\varepsilon_n>0$ for all $n$. Furthermore, this approximation results in exact eigenvalues in case of shape-invariant systems.

Let us now utilise the relation (\ref{3.2}) between the eigenvalues of the Dirac Hamiltonian and the Witten Hamiltonians. For unbroken SUSY the relativistic version of the quasi-classical approximation then reads
\begin{equation}\label{5.3}
  \int_{x_{\rm L}}^{x_{\rm R}}{\rm d} x \sqrt{E^2 - m^2c^4 - W^2(x)} = c\hbar\pi n\,,\qquad n\in\mathbb{N}_0\,,
\end{equation}
where the turning points $x_{\rm L} \leq x_{\rm R}$ are now given by the relation $W^2(x_{\rm L/R})=E^2-m^2c^4$. Again for $n=0$ this results in the exact ground-state energy $E_0^-=-mc^2$. The approximate eigenvalues are given by $E_n^\pm=\pm\sqrt{E^2}$ for $n\in\mathbb{N}$. Similar, for broken SUSY, we have
\begin{equation}\label{5.4}
  \int_{x_{\rm L}}^{x_{\rm R}}{\rm d} x \sqrt{E^2 - m^2c^4 - W^2(x)} = c\hbar\pi \left(n-\frac{1}{2}\right)\,,\qquad E_n^\pm=\pm\sqrt{E^2}\,,\qquad n\in\mathbb{N}\,.
\end{equation}
Again, as in the non-relativistic case, whenever the SUSY potential $W$ is shape-invariant both formulas (\ref{5.3}) and (\ref{5.4}) reproduce the exact spectrum of the corresponding Dirac Hamiltonian. We leave it as an exercise to the reader to verify that for $W(x)=mc\omega x$ the formula (\ref{5.3}) reproduces the spectrum of the Dirac oscillator as given in (\ref{3.24}).

\section{Conclusions}
In the current paper we have studied the most general supersymmetric Dirac Hamiltonian in $(1+1)$ space-time dimensions. It has been shown that the spectral properties of the relativistic Hamiltonian $H_{\rm D}$ as given in eq.\ (\ref{6}) can be reduced to the non-relativistic spectral problem (\ref{13}) of the Witten model. Hence the main results are represented by eq.\ (\ref{3.2}) and eq.\ (\ref{3.17}). In addition, we have shown that the relativistic resolvent kernel (\ref{4.1}) can be expressed in terms of those of the Witten model (\ref{4.5}) via eq.\ (\ref{4.6}). As an explicit example we have chosen the Dirac oscillator for which the closed-form expression (\ref{4.14}) of the Green's function was obtained for the first time. Obviously, any of the shape-invariant non-relativistic SUSY models, which exhibit exact solutions, immediately result in exact solutions of the corresponding relativistic model.

With the discussion of section \ref{sec:6} we may also consider not exactly solvable systems by applying the supersymmetric quasi-classical approximations. Here for example, one could study  anharmonic oscillator systems being characterized by a SUSY potential $\Phi(x)= |x|^d$ and $\Phi(x)={\rm sgn}\, x |x|^d$ exhibiting broken and unbroken supersymmetry, respectively. It shall be noted that in the limit $d\to\infty$ theses systems simulate a particle in the box with various boundary conditions and the quasi-classical approximation is known \cite{Jun2019} to become exact in that limit, too. Hence, this will provide another route towards the study of a Dirac particle in a box, a topic being still of interest \cite{DiracInBox1,DiracInBox2,DiracInBox3,DiracInBox4,DiracInBox5}.

Despite the fact that the current paper focuses on the $(1+1)$ dimensional Dirac systems some of the present results are valid in the more general case of arbitrary supersymmetric Dirac Hamiltonians. For example, the results of section \ref{sec:4} and in particular the explicit expression (\ref{3.14}) for the unitary transformation matrix $U$ is valid for an arbitrary SUSY Hamiltonian (\ref{3}) as long as $M_+=M_-=mc^2$, which is the case for almost all supersymmetric Dirac Hamitonians \cite{Jun2019}.

Finally, let us mention that the present results also apply to radial symmetric Dirac Hamiltonians being supersymmetric. In fact, due the spherical symmetry the radial Dirac Hamiltonian is of the same form as given in (\ref{6}), cf.\ eq.\ (9.102) in \cite{Jun2019},  with a pseudo-scalar potential being of the form $W(r)=\kappa/r+\Phi(r)$ now acting on the positive half-line $r\in\mathbb{R}^+$ and $\kappa$ denotes the eigenvalues of the spin-orbit operator.


%


\end{document}